\documentclass{webofc}
\usepackage[varg]{txfonts}   
\begin{document}
\title{The ups and downs of inferred cosmological lithium}
%
%

\author{\firstname{Andreas} \lastname{Korn}\inst{1}\fnsep\thanks{\email{andreas.korn@physics.uu.se}} 
}

\institute{Division of astronomy and space physics, Department of physics and astronomy, Uppsala University, Uppsala, Sweden 
          }

\abstract{%
I summarize the stellar side of the cosmological lithium problem(s). Evidence from independent studies is accumulating and indicates that stars may very well be fully responsible for lowering their surface lithium from the predicted primordial value to observed levels through internal element-transport mechanisms collectively referred to as atomic diffusion. 

While atomic diffusion can be modelled from first principles, stellar evolution uses a parametrized representation of convection making it impossible to  predict convective-boundary mixing as a vital stellar process moderating atomic diffusion. More work is clearly needed here for a fully quantitative picture of lithium (and metallicity) evolution as stars age. 

Lastly, note that inferred stellar lithium-6 abundances have all but disappeared. 

}
\maketitle
\section{Introduction}
\label{intro}
The cosmological lithium problem has been with us for more than two decades. It is the most glaring discrepancy between predictions of standard Big Bang Nucleosyntesis (BBN) calibrated on Cosmic Microwave Background (CMB) anisotropies as observed by WMAP and Planck and lithium abundances inferred from the spectra of warm halo stars. The discrepancy is claimed to have "resisted decades of attempts by cosmologists, nuclear physicists, and astronomers alike" \cite{2022APS..APRK02001K}.
But is this really the case? I will argue that significant progress in modelling how stars deplete their lithium has been made since the turn of the century. Given this observationally verified understanding, stellar depletion is the best physically motivated solution to the classical cosmological lithium problem (see Sect.~\ref{sec-lithium7}). Furthermore, the tentative detections of lithium-6 in warm halo stars, potentially posing another cosmological lithium problem, have  disappeared with better stellar modelling (see Sect.~\ref{sec-lithium6}). By way of introduction, let us look at how stellar lithium abundances are inferred. 

\section{Observing lithium and inferring its abundance}
\label{sec-observations}
For stars relevant to this discussion (spectral types F, G and K), there is practically only one spectral line that can be observed: the resonance doublet at 6707\,\AA\ (see Fig.\ 1). Observed at sufficiently high resolving power ($R$) and signal-to-noise ratio (SNR), it is markedly asymmetric, as the two components have different line strengths and cannot be resolved under the convective conditions prevalent in the atmospheres of these types of stars. (The isotopic shift places the resonance doublet of lithium-6 within the same spectral feature. It creates an additional line asymmetry in the red wing, similar to that caused by stellar surface convection.)

\begin{figure}[h]
\hspace*{-1cm}
\includegraphics[width=14cm,clip]{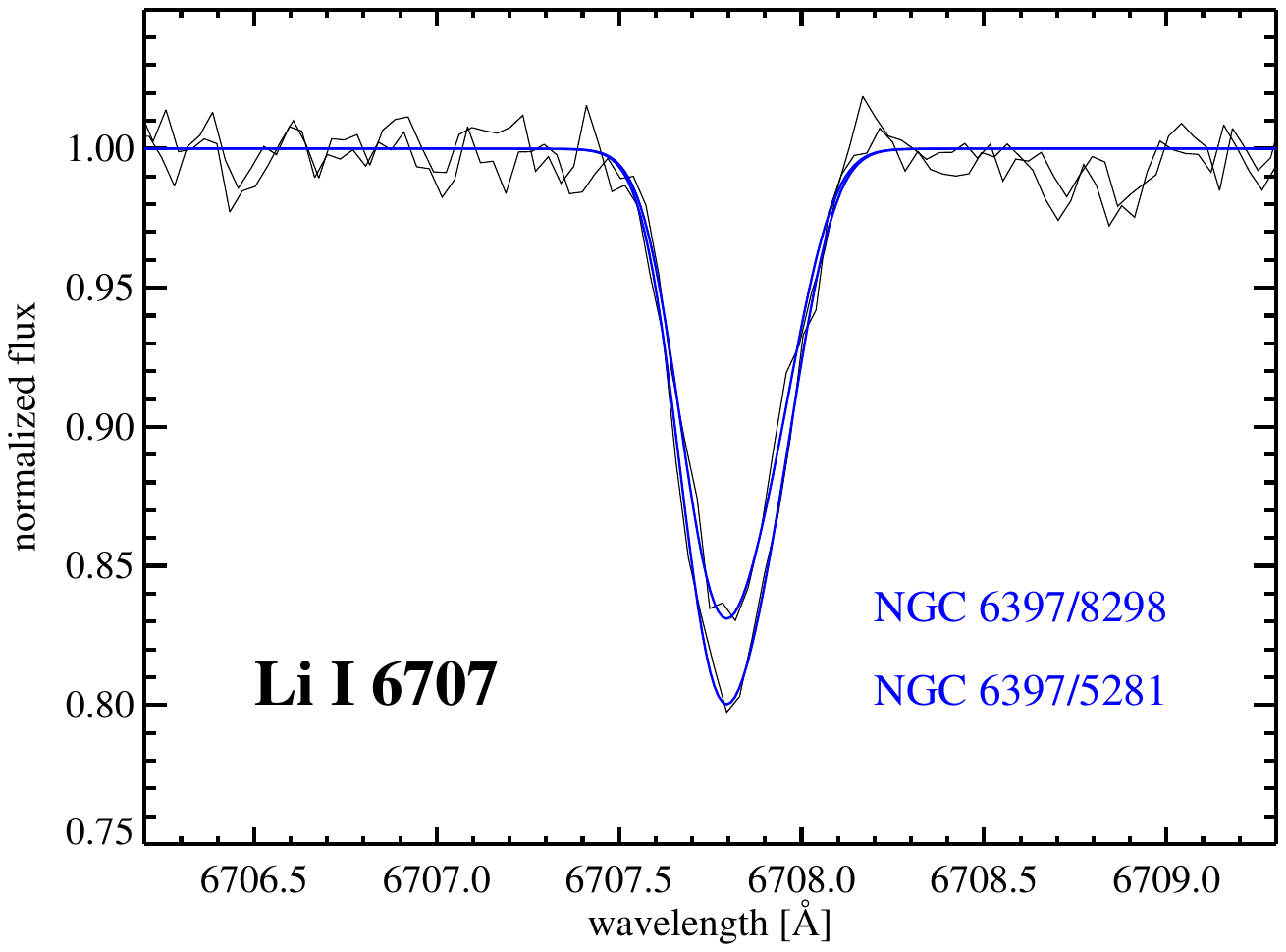}
\vspace*{-15mm}
\caption{The lithium resonance line in two subgiant stars of globular cluster NGC 6397 \cite{2007ApJ...671..402K}. The line is visibly asymmetric, as it is a doublet of two components. One needs good data to measure the lithium abundance from such a line (here SNR > 100 and $R$ > 45,000). Relative to less evolved turnoff-point stars, these two stars show higher abundances of Li and other metals (Mg, Cr, Fe). This is a sign of the interplay of atomic diffusion and convective dredge-up. (\copyright\ AAS. Reproduced with permission)}
\label{fig-1}       
\end{figure}

Once one has a measure of the flux eaten out of the continuum by lithium atoms in the stellar atmosphere, one has to model the line formation to infer the lithium abundance. There are publicly available codes for this, e.g.\ the programme suite SME \cite{2017A&A...597A..16P} to which we are currently adding a server-based frontend (http://webSME.chetec-infra.eu). SME uses classical model atmospheres with mixing-length convection and computes synthetic spectra in local thermodynamic equilibrium (LTE) or non-LTE. The latter approximation is more realistic for lines originating from a transition within a minority species like neutral lithium. Corrections for shortcomings of classical model atmospheres have recently been published \cite{2021MNRAS.500.2159W} and represent the current state-of-the-art in stellar modelling. To the best of our knowledge, the line formation of lithium is now well-understood and free from large systematic biases.  

For several decades, the focus of cosmological-lithium studies was on individual field stars making up the Spite plateau of warm halo stars \cite{1982A&A...115..357S}. Its properties, in particular the absence of detectable star-to-star dispersion (a few outliers aside), was initially hard to reconcile with models predicting the gradual depletion of lithium (and other elements heavier than hydrogen) through processes collectively referred to as atomic diffusion \cite{1984ApJ...282..206M}. So it seemed as if these halo stars had fully preserved the primordial lithium in their atmospheres. However, it could be shown that the Spite-"plateau" value is likely a function of stellar metal content \cite{1999ApJ...523..654R} thus requiring an extrapolation to zero metallicity (like the one applied to derive the primordial helium abundance from star-forming (HII) regions). In other words, the plateau value is not equal to the primordial one!

Qualitative news on primordial vs.\ stellar lithium came from more complete stellar\linebreak models. By introducing a simple parametrized form of additional mixing at the bottom of the convection zone, the Spite plateau predicted from stellar-structure models could be kept thin and flat in the presence of significant (0.4\,dex) stellar-surface depletion (\cite{2005ApJ...619..538R}). A stellar solution through atomic diffusion seemed plausible. But how could one test it?

The key was to study atomic diffusion in stars with a common origin. Nature provides such laboratories in the form of clusters of stars born from the same gas at the same time. By observing stars in different evolutionary stages (from the main-sequence turnoff point, TOP, to the red-giant branch, RGB), atomic diffusion could be studied for a set of heavy elements (mostly Mg, Ti, Ca, Cr and Fe all showing element-specific diffusion trends). The amplitide of the abundance trends between TOP and RGB (0.05-0.25\,dex) dictates how much additional mixing we need for the inferred abundances to agree with those predicted by the atomic-diffusion model. The calibrated model was then applied to lithium \cite{2007ApJ...671..402K,2009A&A...503..545L}. This resulted in good agreement between WMAP-calibrated BBN predictions and inferred birth-cloud lithium abundances of stars in NGC 6397 at a metallicity of [Fe/H]\,=\,$-2.1$ (\cite{2006Natur.442..657K}). A diffusion signature  between turnoff-point stars and subgiants was identified for the first time and can be understood in terms of the interplay of atomic diffusion, the additonal-mixing layer and the expansion of the convective envelope in connection with the first dredge-up \cite{2007ApJ...671..402K}.

We and others have continued studies of this kind and find atomic diffusion at work in both globular and open clusters of various metallicities ranging from [Fe/H]\,=\,$-2.3$ to solar, including the solar-age solar-metallicity cluster Messier 67, see e.g.\ \cite{2019ApJ...874...97S}. It is a physical effect generally operating in such stars, including the sun where diffusion improves the agreement between helioseismology and the standard solar model \cite{2016LRSP...13....2B}. 

Other researchers yet continued to study stars on and around the field-star Spite plateau, see e.g.\ \cite{2010A&A...515L...3M}. These authors could show that there is a mass-dependent amount of lithium depletion well-described by stellar-structure models with atomic diffusion and efficient additional mixing at the bottom of the convection zone. Such high mixing efficiencies are, however, not seen in studies of more evolved stars in metal-poor globular clusters (with the exception of M4 at a metallicity of [Fe/H]\,=\,$-1.1$ \cite{2024MNRAS.527.12120}). 

With respect to a BBN lithium abundance of around log\,$\epsilon$(Li)\,$\simeq$\,2.7 on the customary log\,$\epsilon$(H)\,=\,12 scale \cite{2018PhR...754....1P}, the diffusion-corrected stellar lithium abundances fall systematically short by 0.1-0.15\,dex. This fact may indicate partial destruction of lithium during star formation \cite{2015MNRAS.452.3256F} or a more general redistribution or destruction of lithium in the universe briefly discussed in Sect.\ \ref{sec-misc}. Future observations will likely tell.  

\section{Lithium-6: first off the chart, now off the table}
\label{sec-lithium6}
As early as the 1990, some lithium-6 was claimed to have been found in individual stars on the Spite plateau \cite{1993ApJ...408..262S}. Much later, based on better data a lithium-6 plateau was claimed to exist at the level of a couple of percent of the lithium-7 abundance \cite{2006ApJ...644..229A}. This claim was controversial from the start, because a) the observable effect is subtle and very difficult to quantify, b) was not based on models properly accounting for convective line asymmetries which produce a similar spectral signature as the presence of lithium-6 and c) stellar evolution predicts lithium-6 to be affected by pre-main-sequence destruction. 

Subsequent work with better data and state-of-the-art modelling showed that all previously claimed lithium-6 detections were systematically overestimated \cite{2013A&A...554A..96L}. There may be lithium-6 in metal-rich stars (e.g.\ produced by spallation reaction in the interstellar medium or through stellar activity \cite{2017A&A...604A..44M}), but the local halo field stars used to study primordial lithium have lithium-6/lithium-7 ratios fully compatible with zero \cite{2022MNRAS.509.1521W}. The upper limits have been used to test stellar lithium-7 depletion models concluding that stars can indeed be fully responsible for the cosmological lithium-7 discrepancy \cite{2022JCAP...10..078F}.    

In summary, the stellar verdict for lithium-6 in connection with BBN is the following: There is no lithium-6 plateau among warm halo stars. Consequently, it seems groundless from an observational point of view to discuss lithium-6 in the context of BBN.  

\section{Lithium-7: plateaus abound and remaining offsets}
\label{sec-lithium7}
Additional evidence for stellar physics at work in shaping stellar lithium abundances comes from stars on the lower RGB: these stars display another lithium plateau after surface dilution due to the first dredge-up has taken place \cite{2022A&A...661A.153M}. The lithium abundances are thus lowered to around log\,$\epsilon$(Li)\,$\simeq$\,1. More importantly, this plateau extends to lower metallicities than the Spite plateau which fans out ("melts down" \cite{2010A&A...522A..26S}) at the lowest metallicities making a simple extrapolation to zero metallicity difficult. Even this plateau can be explained by the atomic-diffusion models discussed above. Just like in the case of globular clusters, there seems to be a slight remaining offset to BBN predictions of 0.15\,dex \cite{2022A&A...661A.153M}.   

By digging deep into the Galactic halo, the GALAH survey succeeded in finding a rare group if mildly metal-poor stars which seem to not have undergone lithium depletion. Their lithium abundances, extrapolated to lower metallicities, are fully compatible with BBN predictions \cite{2020MNRAS.497L..30G}. This finding is interpreted in analogy to Population I turnoff-point stars in open clusters: sufficiently hot F-type stars are found to retain their surface lithium, as surface convection is rather superficial and rotational mixing acts to counterbalance atomic diffusion. With upcoming deep and wide stellar surveys like 4MOST, we may find more of these stars populating the unaltered lithium plateau. A mesmerizing prospect! 

\section{Other proposed solutions}
\label{sec-misc}
There are other clever ideas about how to lower stellar surface lithium from a high primordial abundance. 

A fraction of all gas ending up in the potential wells of baryons where we observe stars was undoubtedly cycled through the very first generation of stars, Population III. This process will destroy lithium and lower the amount of lithium to make Population II stars from. By how much is debated. Some authors tried to deplete lithium to the levels observed in Spite-plateau stars \cite{2006ApJ...653..300P}, but this causes problems with overproduction of other elements.  

Lithium may be rare just where we can observe it, because being mostly ionized after recombination it will be preferentially pushed out of potential wells together with H$^+$ \cite{2012arXiv1208.0793P}.  

As for other solutions in particle physics and non-standard cosmology, see the contribution of Carlos Bertulani (these proceedings). Using a Tsallis (rather than the usual Maxwellian) distribution for the velocities of nucleons in BBN may seem like an attractive ansatz to use \cite{2017ApJ...834..165H}, but is at present hardly physically motivated for use in BBN, even if other branches of physics successfully make use of non-extensive statistics. 

In all these cases, one should not turn a blind eye to the stellar solution presented in Sect.~\ref{sec-observations} and \ref{sec-lithium7} modelling physics we know takes place inside stars. If it is confirmed to fall slightly (0.1-0.15\,dex in log(abundance), 25-40\% in linear abundance) short of the most comprehensive and reliable prediction for the primordial lithium abundance, then this remaining gap should be the target for further studies. 

In the context of BBN, lithium has gone from a signpost of new physics to a dissonant detail in an otherwise utterly convincing early-universe model which predicts the primordial abundances from H to Li spanning nine orders of magnitude. BBN abundances set the stage for many problems to solve in 14 billion years of subsequent cosmic (lithium) evolution.

\end{document}